\documentclass[acmtog]{acmart}
\acmSubmissionID{000} 

\usepackage{booktabs} 
\usepackage{multirow}
\usepackage{tabularx}
\usepackage{enumitem}

\usepackage[ruled]{algorithm2e} 

\SetAlFnt{\small}
\SetAlCapFnt{\small}
\SetAlCapNameFnt{\small}
\SetAlCapHSkip{0pt}

\newcolumntype{P}[1]{>{\centering\arraybackslash}p{#1}}

\acmJournal{TOG}

\begin{teaserfigure}
	\centering
	\includegraphics[width=7.1in]{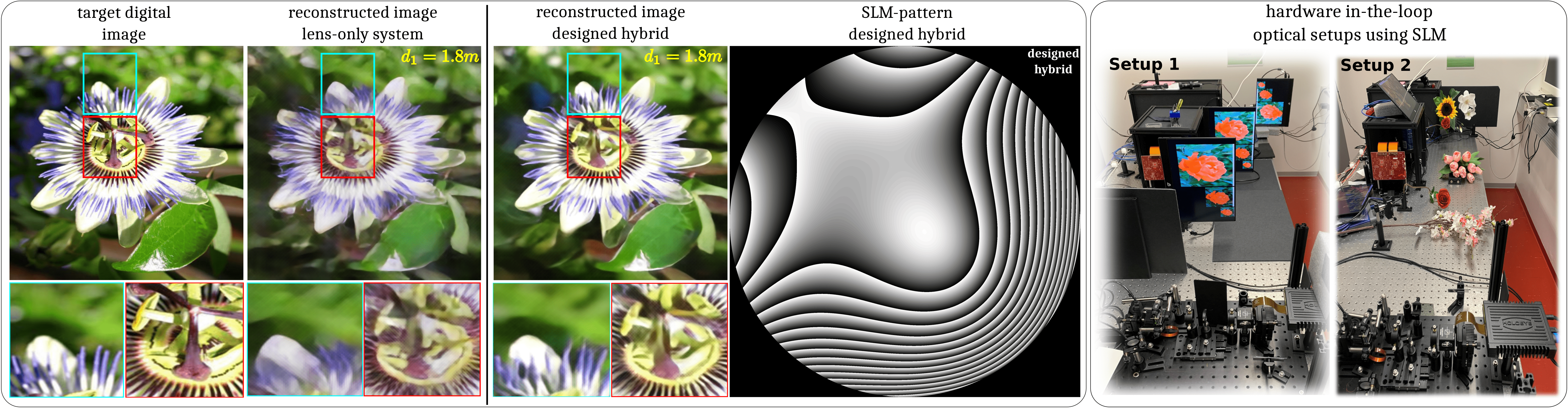}\vspace{-1em}
	\caption{Experimental setups and reconstructed images. The last two images show optical Setups 1 and 2 used for: (1) the system to be optimized and (2) testing setup for all-in-focus imaging. Setup 1 is composed of three monitors displaying images of different distances from the sensor. These images are registered and processed jointly for design purposes and separately for image reconstruction. In Setup 2, the objects (flowers) are shared in the imaging field-of-view, a single image is registered and processed. The SLM phase-pattern demonstrates an optimized profile of the designed encoding phase modulation. The flower images show reconstructions by the designed hybrid optical system (refractive lens and SLM as DOE) and the lens-only system (also with optimized image reconstruction) obtained in Setup 1. The registered blur observations are obtained for the defocus distance 1.8~m.} \vspace{1em}
	\label{fig:initial}\vspace{-1em}
\end{teaserfigure}

\begin{document}
\title{Hybrid Diffractive Optics Design via Hardware-in-the-Loop Methodology for Achromatic Extended-Depth-of-Field Imaging}

\author{Samuel Pinilla}
\orcid{1234-5678-9012-3456}
\affiliation{%
 \institution{University of Manchester at Harwell Science and Innovation campus}
 \city{Didcot}
 \postcode{Oxon OX11 0FA}
 \country{UK}}
\email{samuel.pinilla@manchester.ac.uk}
\author{Seyyed Reza Miri Rostami}
\affiliation{%
 \institution{Tampere University}
 \city{Tampere}
 \country{Finland}
}
\email{SeyyedReza.MiriRostami@tuni.fi}
\author{Igor Shevkunov}
\affiliation{%
 \institution{Tampere University}
 \city{Tampere}
 \country{Finland}}
\email{igor.shevkunov@tuni.fi}
\author{Vladimir Katkovnik}
\affiliation{%
 \institution{Tampere University}
 \city{Tampere}
 \country{Finland}
}
\email{vladimir.katkovnik@tuni.fi}
\author{Karen Eguiazarian}
\affiliation{%
 \institution{Tampere University}
 \city{Tampere}
 \country{Finland}}
\email{karen.eguiazarian@tuni.fi}


\begin{abstract}
End-to-end optimization of diffractive optical elements (DOEs) profile through a digital differentiable model combined with computational imaging have gained an increasing attention in emerging applications due to the compactness of resultant physical setups. Despite recent works have shown the potential of this methodology to design optics, its performance in physical setups is still limited and affected by manufacturing artifacts of DOE, mismatch between simulated and resultant experimental point spread functions, and calibration errors. Additionally, the computational burden of the digital differentiable model to effectively design the DOE is increasing, thus limiting the size of the DOE that can be designed. To overcome the above mentioned limitations, the broadband imaging system with phase-only spatial light modulator (SLM) as an encoded diffractive optical element is proposed and developed in this paper. The SLM is exploited as a pixel-wise programmable device for design of a light-modulation phase pattern. A co-design of the SLM phase pattern and image reconstruction algorithm is produced following the end-to-end strategy, using for optimization a convolutional neural network equipped with quantitative and qualitative loss functions. The optics of the imaging system is hybrid consisting of SLM as DOE and refractive lens. SLM phase-pattern is optimized by applying the Hardware-in-the-loop technique, which helps to eliminate the mismatch between numerical modeling and physical reality of image formation as light propagation is not numerically modeled but is physically done. In our experiments, the hybrid optics is implemented by the optical projection of the SLM phase-pattern on a lens plane for a depth range 0.4-1.9~m (Figure \ref{fig:initial}). Multiple numerical and physical experiments confirm high-quality imaging of the system in the achromatic extended depth-of-field scenario. Comparison with compound multi-lens optics such as Sony A7 III and iPhone Xs Max cameras show that the proposed system is advanced in all-in-focus sharp imaging.
\end{abstract}

\ccsdesc[100]{Hardware-in-the-loop design}
\ccsdesc[500]{Hybrid diffractive optics}
\ccsdesc[300]{Achromatic extended-depth-of-field}

%
%

\keywords{Hardware-in-the-loop for design of DOE phase pattern}

\maketitle

\section{Introduction}
\label{sec:introduction}
Computational imaging with encoded diffractive optical elements (DOEs) (e.g. binary-, multi-level phase elements) and meta-optical elements (MOEs) is a multidisciplinary research field in the intersection of optics, mathematics and digital image processing \cite{10.1145/3197517.3201333,chang2018hybrid,Dun:20,arguello2021shift,colburn2021inverse}. It is based on a combination of optical encoding (at the optics layer) and algorithmic decoding (at the image processing layer). Contrary to the traditional optical systems with refractive lenses, the hardware decoding (optical focusing) is replaced by the software computational decoding \cite{antipa2018diffusercam,7517296,7559956,Mosleh_2020_CVPR}. In DOE, this phase coding is induced due to the length of the ray path inside a DOE material. In MOE, the phase delay is induced via the response of nanostructures (so-called nanoantennas) built on the surface of the substrate material \cite{engelberg2020advantages}. Both DOEs and MOEs are used as the wavefront coding instruments \cite{chen2020flat}. Despite the fundamental difference in nature between DOEs and MOEs, some authors treat MOEs as a special class of DOEs, and we also use DOE as a notation for both classes of these elements.

The potential of DOEs as optical elements is three-fold: (1) Accuracy and quality of imaging can be high at least on the level of the compound conventional refractive optics, while DOEs are compact (thickness in micrometers), light, and cheap. (2) The systems with DOEs have the potential to allow solutions beyond the ability of the conventional systems, in particular, for some pattern recognition problems, hyperspectral imaging, and even in such conventional types of problems as extended depth of field (EDoF) \cite{10.1117/1.OE.60.5.051204}, extended field-of-view, and achromatic broadband imaging. Typical examples are flat hyperspectral cameras \cite{Monakhova:20} and flat face recognition devices \cite{8590781}. (3) Nearly arbitrary manipulations of wavefields are possible due to amazing progress in nanotechnology. While nanotechnology allows obtaining DOEs with nearly arbitrary phase-patterns, the desirable optical characteristics as a rule are a priori unknown and an ordinary intuition usually cannot help, what makes the design of encoded DOEs quite problematic. Images registered by sensors in systems with encoded DOEs are blurred (assumed to be convolution between scene and a unique point spread function (PSF)). Sometimes they are strongly blurred and even completely unrecognizable, and the ﬁnal high-quality sharp images are achieved only after computational inverse imaging \cite{Tseng2021NeuralNanoOptics}. The today well established practice is to formalize the deal and to obtain these characteristics as well as DOE phase profiles from solutions of end-to-end optimization problems (e.g. \cite{baek2020end,sun2021end,liu2021end}).

However, when a found desirable encoded DOE leaps from paper (theory) to practice, a natural concern is how well the physical implementation of the DOE preserves the desirable mathematical model and its advisable performance. Specifically, this gap between theoretical solution and real-life implementation comes from the limited performance in physical setups due to disturbing artifacts in DOE implementation, a mismatch between simulated and resultant experimental PSFs, calibration errors, etc.

\begin{figure}[t!]
	\centering
	\includegraphics[width=1.0\linewidth]{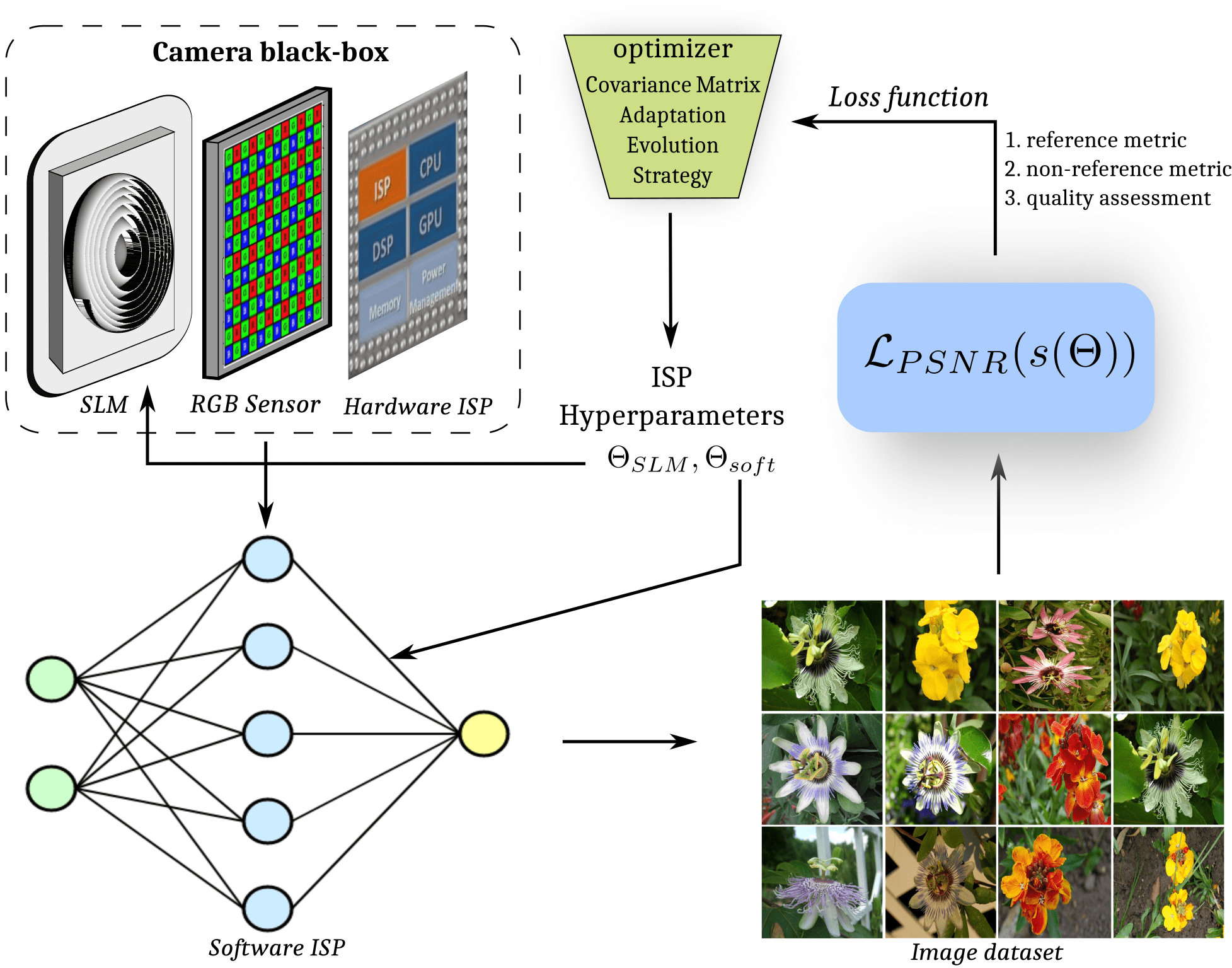}\vspace{-1em}
	\caption{\small Schematic of the proposed HIL setup and optimization of SLM program (SLM) and Software ISP. The camera black-box model is composed of SLM (varying), sensor and hardware ISP (fixed). The output of the latter is an input of the designed image processing algorithm (Software ISP).}\vspace{-2em}
	\label{fig:hardwareLoop}
\end{figure}

Bridging the gap between theoretical solution and real-life physical implementation is the primary goal of this work. Therefore, in this paper, we propose an end-to-end design of the DOE through a ‘Hardware-In-the Loop (HIL)’ imaging setup, for achromatic EDoF, using a programmable phase Spatial Light Modulator (SLM) for implementation of encoded DOEs as illustrated in Figures \ref{fig:initial} and \ref{fig:hardwareLoop}. The hyperparameter of SLM, $\Theta_{SLM}$, and image signal processing (ISP) algorithms, $\Theta_{soft}$, are obtained due to end-to-end optimization fitting reconstructed and true images. Fundamentally, our HIL-SLM setup guarantees a proper 'modeling' of the black-box hardware in optimization without all discrepancies between mathematical models and physical reality typical for the state-of-the-art model-based design approaches. Especially important, that all errors in modeling of wavefront propagation through DOEs (difficult to model mathematically and numerically) will be taken out due to optimization produced for real physical use of the hardware.

For design and optimization of SLM phase-pattern, we use the Covariance Matrix Adaptation Evolution Strategy (CMA-ES) optimizer \cite{auger2012tutorial,dufosse2021augmented}, which is a zero-order algorithm, following a parameterized model of SLM phase-pattern as a linear combination of the first fourteen Zernike polynomials where coefficients are summarized in $\Theta_{SLM}$. The crucial advantage of CMA-ES for the HIL setup is that it does not require the knowledge of the gradient to optimize the optics. The phase-pattern of SLM is designed as a piece-wise invariant Multilevel Phase-Pattern (MPP) defined for the design wavelength. The software ISP is designed using CNNs as a natural tool for optimization of image reconstruction algorithms with respect to the hyperparameter $\Theta_{soft}$ using Adam algorithm \cite{kingma2014adam}.

We assume that the hardware ISP, in particular including demosaicing, is fixed. Other components of the hardware (sensor and hardware ISP: low-level image processing present in most digital cameras) also can be a subject of optimization provided that they are variable. Together with SLM and sensor, it outlines a camera as a ‘black-box’ (given as an input-output system of an unknown mathematical model) embedded in algorithmic optimization. Considering the advantages of our HIL setup we have designed a DOE of $9.2mm$ diameter with a pixel resolution of $3.74\mu m$ and $10mm$ focal length. To the best of our knowledge these DOE parameters have not been achieved using state-of-the-art digital differentiable models due to limitations on computational costs (e.g. \cite{Dun:20}).

In experiments for all-in-focus imaging, we compare the designed system with conventional compound multi-lens cameras such as iPhone Xs Max and Sony A7 III with the objective lens configured with 85mm focal length and F22. For the Sony camera, we chose the largest F-number (corresponding to the smallest aperture size) to result in the deepest DoF. Our system with hybrid optics designed in HIL strategy provides performance that is competitive in terms of imaging quality and advanced achromatic sharp imaging.

The contribution of this work can be itemized as follows:
\begin{itemize}[leftmargin=*]
 \item Broadband diffractive imaging with programmable SLM phase pattern projected to the lens plane with online physical tuning of DOE (for the first time); 
 \item The framework for HIL co-design methodology of multilevel SLM phase-pattern (MPP) and inverse imaging for achromatic EDoF;
 \item Advanced performance in terms of EDoF for the developed imaging system in comparison with compound optics of SONY A7 III (85mm/F22.0) and iPhone Xs Max;
\end{itemize}

\textit{Scope.} We believe that the proposed end-to-end HIL design framework provides steps forward to mitigate effect of uncontrolled artifacts and mismatch between mathematical modeling and real-life experiments. It is worth mentioning that programmable phase-only SLMs are not compact, light and cheap, contrary to it, they are bulky and very expensive. However, the designed optimal SLM phase-pattern can be used as desirable phase profile for manufacturing of hardware DOEs such as diffractive and meta optical elements.\vspace{-0.5em}

\section{Related Work}
\label{Related Work}
As relevant to this work, we provide references to achromatic EDoF imaging with DOEs and the hardware-in-the-loop technique as applied to imaging problems.

\begin{figure}[t!]
	\centering
	\includegraphics[width=1.0\linewidth]{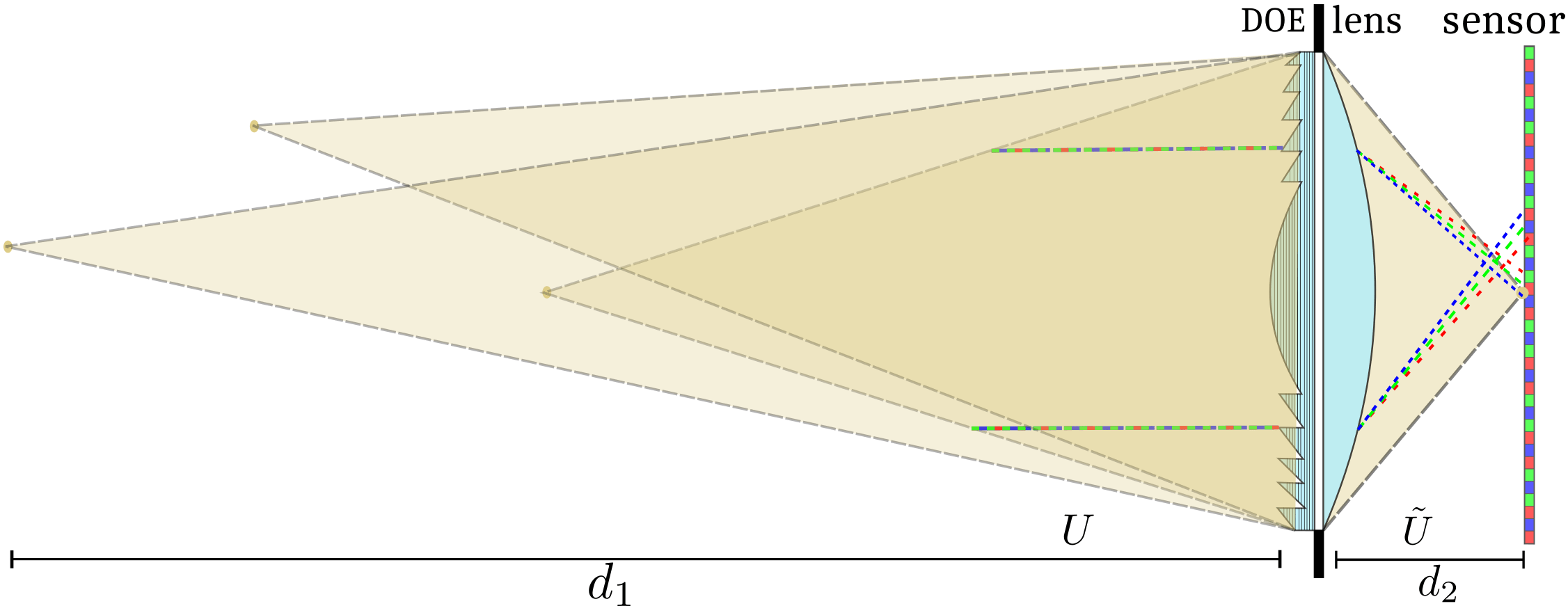}\vspace{-1em}
	\caption{\small Hybrid system for extended-depth-of-field imaging. Light wave from different distances $d_1$ propagates on the aperture plane containing DOE to be designed. The DOE modulates the phase of the incident wavefront. The resulting wavefront propagates through the lens to the aperture-sensor, distance $d_2$.}\vspace{-0.5em}
	\label{fig:scheme}
\end{figure}

\textit{Achromatic EDoF imaging.} EDoF with minimal chromatic aberrations (see Figure \ref{fig:scheme}) is a classic problem of computational imaging that has been intensively studied in recent works motivated by development of new lighter, and compact optical systems for emerging imaging applications such as computational photography. The reason to study this problem in modern times is because classical compound multi-array of lenses solutions do not either apply \cite{Tseng2021NeuralNanoOptics} or extend the DoF for the new compact optical systems.

One of the basic tools of all-focus imaging is an optics with depth invariant Point Spread Function (PSF) referring to the prominent wavefront coding (WFC) proposed in \cite{Dowski:95}. A flow of publications exploiting this idea in different optical setups and for various goals. The recent works develop this idea for imaging with diffractive \cite{10.1145/3197517.3201333} and meta-plates \cite{Tseng2021NeuralNanoOptics,BayatiPestourieColburnLinJohnsonMajumdar+2021,doi:10.1126/sciadv.aar2114}. The designed depth invariant PSFs serve as the main components of inverse imaging. For inverse imaging in physical experiments, the calculated PSFs are calibrated and corrected according to the corresponding test for manufactured DOEs. In this way, a gap between theoretical and physical image formation can be diminished following the logic: first, design of DOE and, after, its tuning according to the experimental study of the manufactured DOEs.

The sensitivity of inverse imaging with respect to errors in PSF design and calculation can be compensated to some extend by using convolutional neural network (CNN) technologies \cite{9546648} dealing directly with the blurred images and designing the inverse imaging algorithms based on test-images as it is particularly done in the works \cite{sun2021end,Tseng2021NeuralNanoOptics,alghamdi2021transfer,liu2021end,chen2020lensless}. Note that in our approach based on HIL-SLM setup this problem is eliminated completely, since we do not deal with PSFs at all as we do not need them for SLM phase-pattern design and do not use them for inverse imaging.

In this work the elements of interest to be jointly designed are refractive lens and DOE (hybrid optics) as it is shown in Figure \ref{fig:scheme} to improve DoF and reduce chromatic aberrations of a system. Hybrid optics for achromatic EDoF appeared in the works \cite{flores2004achromatic,liu2007diffractive,10.1117/1.2430506} where it was designed as a focusing lens with direct evaluation of optical performance. In \cite{Rostami:21}, design in terms of end-to-end optimization was developed targeting phase-encoded inverse imaging with modification of the hybrid to optimal sharing of optical power between DOE and lens.

\textit{Hardware-in-the-loop.} The HIL approach to optimization is known well and for a long time in engineering and science. Actually, any adjustment of hardware when we change something and then look at the result can be treated as the HIL procedure. The main problem is that parameters to be optimized should be easy for variations. 

We know only a few works concerning application of HIL for imaging problems close to being considered in this paper. In the work \cite{peng2020neural}, HIP approach is used to fit parameterized wave propagation model using the phase-only SLM, which results in a dramatically improved performance of the holographic display. In \cite{TYY19}, HIP setup is used for tuning of ISP parameter provided fixed optics. To the best of our knowledge, what is done in this paper is a first attempt to use HIL in design of DOEs for intensity imaging.

\begin{algorithm}[ht]
 \caption{HIL design of SLM phase-pattern and inverse imaging}
 \label{alg:design}
 \SetKwInOut{Input}{inputs}
 \SetKwInOut{Output}{output}
 \SetKwProg{FindAnMFS}{HIL-design}{}{}
 \FindAnMFS{$(\lambda,N_{iter})$}{
 \textbf{Require: }$\Theta_{SLM}^{(0)}$\;
 
 Initialize CMA-ES, $\Theta_{SLM} \leftarrow \Theta^{(0)}_{SLM},t\leftarrow 1$\; 
 Train initial inverse imaging CNN for $0.5,1.0,1.8$m, $\Theta_{soft} \leftarrow \Theta^{(0)}_{soft}$ for a wide range of hyperparameters $\Theta_{SLM}$\;
 \While{$t\leq N_{iter}$}{
 \For{$r=1$ to $R$}{
 $\Theta_{SLM}^{(r)}\leftarrow $ randomly draw from Gaussian at $\Theta_{SLM}$\;
 Add random noise to $\Theta_{SLM}^{(r)}$\;
 $s\left( \Theta_{SLM}^{(r)}\right) \leftarrow $ get blurred data at $0.5,1.0,1.8$m of images $\boldsymbol{I}_{1},\dots, \boldsymbol{I}_{\mathcal{J}}$ and use $\Theta_{soft}^{(t)}$-CNN to estimate them\;
 $\mathcal{L}_{HIL}\left(s\left( \Theta_{SLM}^{(r)}\right)\right) \leftarrow $ compute average PSNR among the estimated $\mathcal{J}$-images at each distance $0.5,1.0,1.8$m\;
 $\Theta_{SLM} \leftarrow $ update CMA-ES\;
 }
 $t \leftarrow t + 1$\;
 $\Theta_{soft}^{(t)} \leftarrow $ train inverse imaging CNN for best SLM-pattern among $\{\Theta_{SLM}^{(1)},\dots, \Theta_{SLM}^{(\lambda)}\}$\;
 }
 \KwRet{$\Theta_{SLM},\Theta_{soft}$}\;
 }
\end{algorithm}\vspace{-1em}

\begin{figure*}[t!]
	\centering
	\includegraphics[width=1\linewidth]{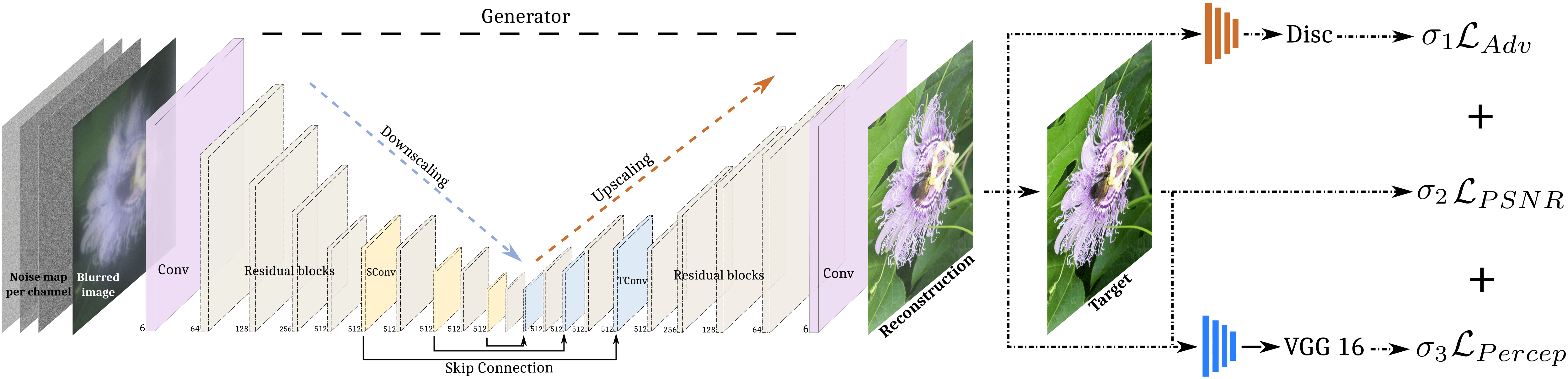}\vspace{-1em}
	\caption{\small Inverse imaging UNet-based neural network for the design of SLM, optimization on $\Theta_{soft}$. The generator model is a U-net architecture that has seven scales with six consecutive downsampling and upsampling operations. We adopt a weighted combination of PSNR between estimated and ground truth images, $\mathcal{L}_{PSNR}$, and perceptual losses $\mathcal{L}_{Adv}$ and $\mathcal{L}_{Percep}$, with weights $\sigma_{1},\sigma_{2}$, and $\sigma_{3}$.}\vspace{-0.5em}
	\label{fig:unet}
\end{figure*}

\section{Optimization-in-the-loop of Hybrid Optics}
The proposed design framework of hybrid optics for achromatic EDoF is summarized in Algorithm \ref{alg:design} which follows an alternating methodology: fixing hyperparameter $\Theta_{SLM}$, solving $\Theta_{soft}$ for inverse imaging, updating $\Theta_{SLM}$ and so forth. We implement this methodology because it is faster than training the CNN-based inverse imaging per SLM-pattern iterations. This alternating process starts by randomly selecting an SLM-pattern $\Theta_{SLM}^{(0)}$ and a pre-trained inverse imaging CNN for a wide range of hyperparameter $\Theta_{SLM}$. After this, algorithm acquires a set of blurred images at distances from sensor $0.5,1.0,1.8$~m using the ISP of the optical system, $s(\Theta^{(r)}_{SLM})$, which is then passed to the downstream reconstruction (deblurring using trained CNN) module. The output of the task module is evaluated by domains-specific evaluation metric which in this case is the peak-signal-to-noise-ratio (PSNR), $\mathcal{L}_{HIL}(s(\Theta^{(r)}_{SLM}))$. Then, using the $0th$-order stochastic evolutionary search method CMA-ES\footnote{documentation in python of the CMA-ES optimizer in \url{https://pypi.org/project/cma/}.} \cite{hansen1996adapting}, Algorithm \ref{alg:design} updates $\Theta_{SLM}$ taking advantage of the tested SLM-patterns during the $R$ iterations. Once $\Theta_{SLM}$ is updated, Algorithm \ref{alg:design} refines the CNN-based inverse imaging by training it for the best $\Theta^{(r)}_{SLM}$. Performing the previous alternating process $N_{iter}$ times algorithm returns the updated $\Theta_{SLM}$, and $\Theta_{soft}$. The structure of CNN developed for inverse imaging (optimization on $\Theta_{soft}$) is shown in Figure~\ref{fig:unet}. Algorithm \ref{alg:design} can be also initiated by solutions obtained according to the model-based approach from \cite{Rostami:21}. The number of 'global' iterations of this algorithm for $\Theta_{soft}$ is $N_{iter}$ and $R$ is the number of 'local' iterations for $\Theta_{SLM}$. In this work we fixed $N_{iter}=3$ and $R=500$. In the following sections, more details per each stage are described.\vspace{-0.8em}

\subsection{Parameterized Optics Model}
The phase-profile of SLM is designed as a piece-wise spatially invariant function defined for the design wavelength $\lambda_{0}$. Following to \cite{10.1117/1.OE.60.5.051204}, we start from the absolute phase model which further is wrapped to the interval defined by the modulation range of SLM and discretized with a number of the parameters to be optimized equal to fourteen which we introduce in brief. The proposed absolute phase $\varphi_{\lambda_{0}}$ takes the form
\begin{align}
\varphi_{\lambda _{0}}(x,y) = \frac{2\pi \alpha}{\lambda_{0} f_{\lambda_{0}}} (x^{2} + y^{2}) + \beta(x^{3} + y^{3}) +\sum_{p=1,p\not = 4}^{P}\rho_{p}Z_{p}(x,y),
\label{abs=phase}
\end{align}
where the first term models the squared phase of the lens which fits the form of the fourth Zernike polynomial omitted in the sum of the Zernike components, $\alpha$ encapsulates the focusing contribution of the SLM for wavelength $\lambda _{0}$, and the lens focal length $f_{\lambda_{0}}$ fixed as $f_{\lambda_{0}}=0.01 m$ in this work. The cubic phase of magnitude $\beta$ is a typical component for EDoF as introduced in \cite{Dowski:95}, the third group of the items is for parametric approximation of the free-shape DOE using the Zernike polynomials $Z_{p}(x,y)$ with coefficients $\rho_{p}$ to be estimated. Thus, the full set of parameters to be optimized for SLM is defined as $\Theta_{SLM}=(\alpha, \beta, \rho_{1},\dots,\rho_{P})$.

\textit{Fresnel Order (thickness).} The SLM phase profile in radians is defined as $Q=2\pi m_{Q}$, where $m_{Q}$ is called 'Fresnel order' of the mask which in general is not necessarily integer. Then the phase profile of SLM considering the thickness is calculated as
\begin{equation}
\hat{\varphi}_{\lambda _{0}}(x,y) = mod(\varphi_{\lambda _{0}}(x,y) + Q/2,Q)-Q/2. 
\label{lens4}
\end{equation}
The operation in \eqref{lens4} returns $\hat{\varphi}_{\lambda _{0}}(x,y)$ taking the values in the interval $[-Q/2$, $Q/2)$. The parameter $m_{Q}$ is known as 'Fresnel order' of phase-pattern. For $m_{Q}=1$ this restriction to the interval $[-\pi $, $\pi )$ corresponds to the standard phase wrapping operation. For the experiments we will present in next section $m_{Q}$ is fixed to $1.8$ providing as the maximum phase thickness that our SLM supports e.g. phase interval $[-1.8\pi $, $1.8\pi )$.

\textit{Piecewise Invariant Phase. } The SLM-pattern is defined on $2D$ grid $(X,Y)$ with the computational sampling period (computational pixel) $\Delta _{comp}$. We obtain a piece-wise invariant surface for SLM after non-linear transformation of the absolute phase. The discrete uniform grid of the wrap phase profile $\hat{\varphi}_{\lambda _{0}}(x,y)$ to the $N$ levels is given as $\theta_{\lambda _{0}}(x,y) =\lfloor \hat{\varphi}_{\lambda _{0}}(x,y) /N \rfloor \cdot N$, where $\lfloor w \rfloor$ stays for the integer part of $w$. The values of $\theta_{\lambda _{0}}(x,y)$ are restricted to the interval $[-Q/2$, $Q/2)$. $Q$ is an upper bound for thickness phase of $\theta_{\lambda _{0}}(x,y)$.

The physical size of the SLM's pixel is $m_{w}.$ The computational pixel $\Delta _{comp}$ is naturally larger than $m_{w}$ and can be written as $\Delta _{comp}=Km_{w}$, $K$ is a natural number. $\Delta _{comp}$ serves in this design as a lower bound for size of steps (invariant elements of phase-pattern). Larger $\Delta _{comp}$ means a simpler structure of phase-pattern and a simpler corresponding phase mask for implementation of this phase modulation. Fresnel order $Q$, number of steps $N$ and computational pixel $\Delta _{comp}$ could be included in $\Theta_{SLM}$ as design parameters.\vspace{-0.8em}

\begin{figure}[t!]
	\centering
	\includegraphics[width=1\linewidth]{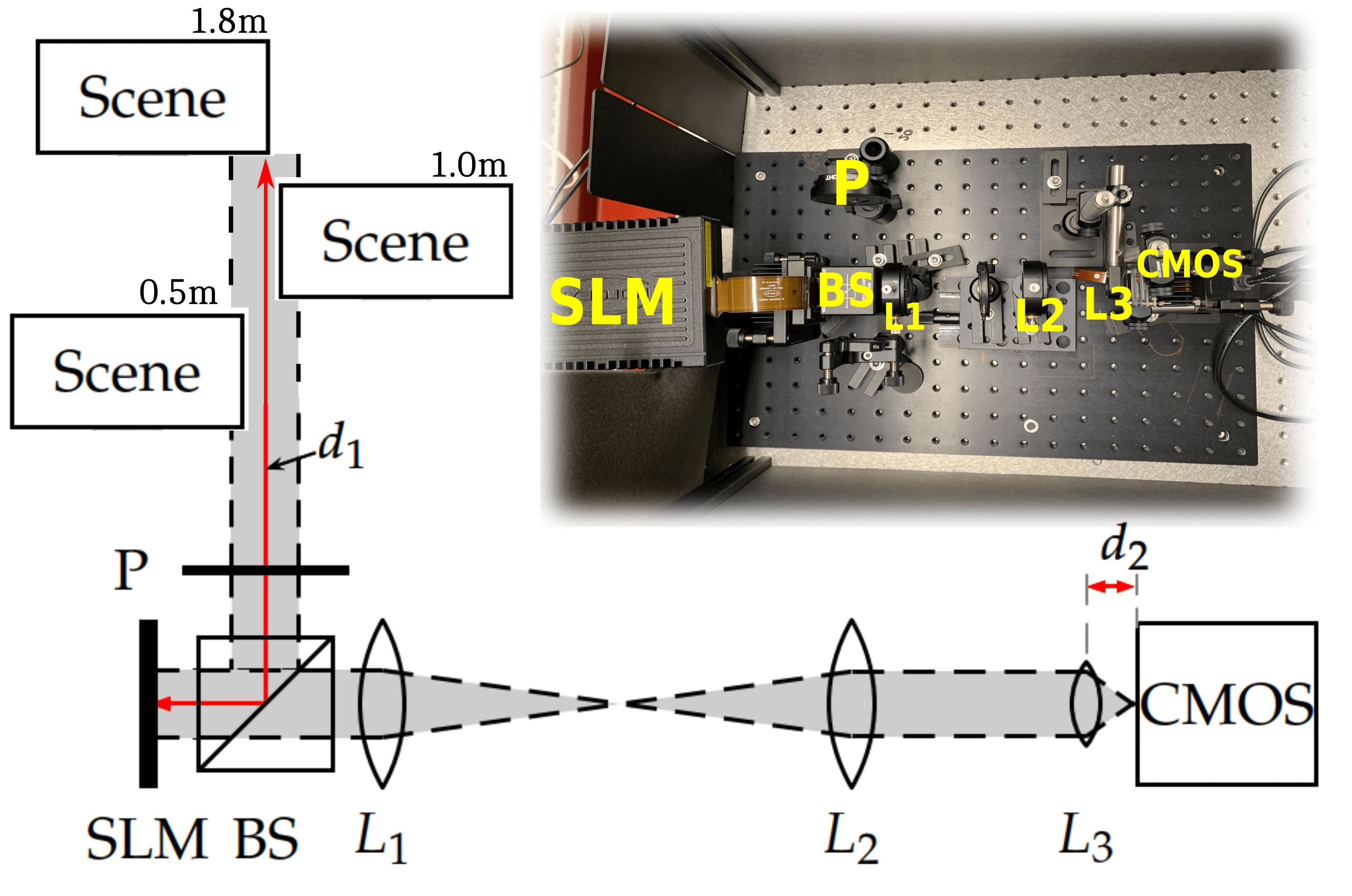}\vspace{-1em}
	\caption{\small Experimental setup. P is a polarizer, BS is a beamsplitter, SLM is a spatial light modulator. The lenses $L_1$ and $L_2$ form the 4f-telescopic system projecting wavefront from the SLM plane to the imaging lens $L_3$, CMOS is a registering camera. $d_1$ is the distance between the scene and the plane of the hybrid optics, and ~$d_2$ is a distance between the optics and the sensor.}\vspace{-1.5em}
	\label{fig:scheme_3}
\end{figure}

\begin{figure*}[t!]
	\centering
	\includegraphics[width=1\linewidth]{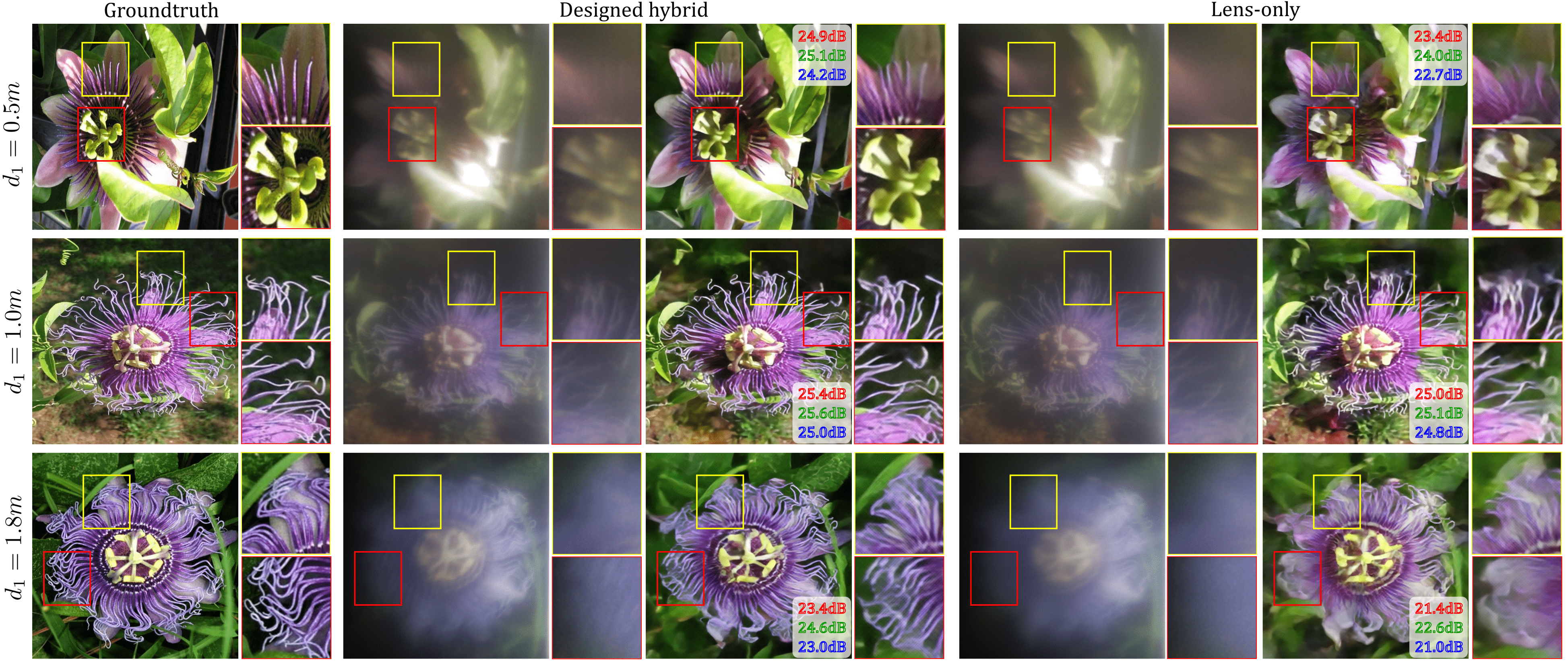}\vspace{-1em}
	\caption{\small Results for Setup 1. The reconstructed images with two zoomed regions at three different distances (imaging monitor-SLM): $d_{1}=0.5,1.0, 1.8$~m, for designed hybrid and lens-only systems. The PSNR values are reported for each depth and each color channel separately. The high-quality imaging with PSNR values of about 25 dB for different imaging depths and colors is achieved by the designed hybrid. In contrast, while the lens-only system performs well for the infocus distance $d_{1}=1.0$~m and the imaging quality degrade essentially for the off-focus distances.} \vspace{-0.5em}
	\label{fig:system}
\end{figure*} 

\subsection{Extended-depth-of-field Imaging}
The optical setup implemented in this work to design a DOE using SLM following a hardware in-the-loop strategy is depicted in Figure~\ref{fig:scheme_3} for the achromatic EDoF problem as illustrated in Figure \ref{fig:scheme}, where 'Scene' denotes objects under investigation; the polarizer, 'P', keeps the light polarization needed for a proper wavefront modulation by SLM; the beamsplitter, 'BS', governs SLM illumination and further light passing; the lenses '$L_1$' and '$L_2$' form a 4f-telescopic system transferring the light wavefront modified by SLM to the lens '$L_3$' plane; the lens '$L_3$' forms an image of the 'scene' on the imaging detector, 'CMOS'. Since the main goal of this paper is to extend the DoF of a photography camera we have three scenes located at different depths which are physically implemented by three identical polarized 15.6'' monitors\footnote{Blackstorm MobileMonitor. Their Description can be found in \url{https://www.verkkokauppa.com/fi/product/14167/mxvvq/Blackstorm-MobileMonitor-15-6-Full-HD-kannettava-naytto.}}.

Test images in 'scene' plane are displayed on these monitors with $1920\times1080$ pixels and $570$~ppi. The distances $d_{1}$ for the monitors are $0.5,1.0,1.8$~m. The image dataset we use as scenes consists of 1000 high-resolution images\footnote{Image databases employed in the training stage can be found in \url{https://dataverse.harvard.edu/dataset.xhtml?persistentId=doi:10.7910/DVN/1ECTVN}.}. We point out that the image sizes at each monitor (depths $0.5,1.0,1.8$~m) are scaled in such a way that the sizes of the corresponding registered images at the sensor are the same, in sensor pixel $512\times 512$.

For DOE implementation, we use the Holoeye phase-only GAEA-2-vis SLM panel, resolution $4160\times2464$, pixel size $3.74~\mu$m; '$L_1$' and '$L_2$' achromatic doublet lenses with diameter $12.7$~mm and focal distance of $50$~mm; BK7 glass lens '$L_3$' with diameter $9.2$~mm and focal distance $10.0$~mm (approximately $d_{2}$); 'CMOS' Blackfly S board Level camera with the color pixel matrix Sony IMX264, $3.45~\mu$m pixels and $2448\times2048$ pixels. SLM allows us to study the hybrid optics with the phase distribution of the designed DOE (implemented on SLM) additive to the imaging lens ‘$L_3$’. The DOE phase was created as an 8-bit \textit{*.bmp} file and imaged on SLM. We calibrated the SLM phase delay response to the maximum value of $3.6\pi$ (fresnel order equal to 1.8) for a wavelength of $510$~nm. This $3.6\pi$ corresponds to the value 255 of \textit{*.bmp} file for the phase image of DOE.\vspace{-0.8em}

\subsection{Inverse Imaging: Update for $\Theta_{soft}$}
Data processing here solves the following problems: 1) inverse imaging (reconstruction of sharp images from the registered blurred ones); 2) compensation of various  errors such as  phase modulation errors in SLM, errors in demosaicing software, noise in CMOS, etc. The learning NN approach within an end-to-end design is used to resolve these problems. In this work we implement a DRUNet CNN architecture \cite{zhang2021plug} illustrated in Figure \ref{fig:unet}. We remark that this network has the ability to handle various noise levels for an RGB image, per channel, via a single model. The backbone of DRUNet is U-Net which consists of four scales. Each scale has an identity skip connection between $2\times 2$ strided convolution (SConv) downscaling and $2\times 2$ transposed convolution (TConv) upscaling operations. The number of channels in each layer from the first scale to the fourth scale are 64, 128, 256 and 512, respectively. Four successive residual blocks are adopted in the downscaling and upscaling of each scale. Each residual block only contains one ReLU activation function. It is worth noting that the proposed DRUNet is biasfree, which means no bias is used in all the Conv, SConv and TConv layers \cite{zhang2021plug}.

An appropriate loss function is required to optimize our inverse imaging to provide the desired output. Thus, we use a weighted combination of PSNR between estimated and ground truth images, $\mathcal{L}_{PSNR}$, and perceptual losses given below by:

\textbf{Perceptual loss: }To measure the semantic difference between the estimated output and the ground truth, we use a pretrained VGG-16 \cite{simonyan2014very} model for our perceptual loss \cite{khan2020flatnet}. We extract feature maps between the second convolution (after activation) and second max pool layers $\varphi_{22}$, and between the third convolution (after activation) and the fourth max pool layers $\varphi_{43}$. Then, the loss $\mathcal{L}_{Percep}$ is the averaged PSNR between the outputs of these two activation functions for both estimated and ground truth images.

\textbf{Adversarial loss: } Adversarial loss \cite{goodfellow2014generative} was added to further bring the distribution of the reconstructed output close to those of the real images. Given the swish activation function \cite{ramachandran2017searching} as our discriminator $D$, this loss is given as $\mathcal{L}_{Adv} = -\log(D(I_{est}))$ where $I_{est}$ models the estimated image.

Our total loss for the proposed CNN inverse imaging while training is a weighted combination of the three losses and is given as, $\mathcal{L}_{CNN}=\sigma_{1}\mathcal{L}_{PSNR} + \sigma_{2}\mathcal{L}_{Percep} + \sigma_{3}\mathcal{L}_{Adv}$, where, $\sigma_{1},\sigma_{2}$ and $\sigma_{3}$ are empirical weights assigned to each loss. In this work these constant are fixed as $\sigma_{1}=1.0, \sigma_{2}=0.6$, and $\sigma_{3}=0.1$. Lastly, the parameters of this networks to be optimized are summarized in $\Theta_{soft}$.\vspace{-1em}

\subsection{HIL Optimizer: Update for $\Theta_{SLM}$}
We propose a nonlinear black-box optimizer that uses CMA-ES strategy in Algorithm \ref{alg:design} to optimize the hyperparameter $\Theta_{SLM}$. The loss function to be optimized by CMA-ES is the averaged PSNR between target image and estimates from CNN-based inverse imaging for each distance $0.5,1.0,1.8$~m, $\mathcal{L}_{HIL}(s(\Theta^{(r)}_{SLM}))$ in Algorithm \ref{alg:design}. Despite the fact that CMA-ES is a $0th$-order stochastic evolutionary search method, it can be viewed as $2nd$-order since it estimates a covariance matrix closely related to the inverse Hessian \cite{hansen1996adapting}. This feature allows CMA-ES to handle badly conditioned problems. In fact, methods that require derivatives like Adam \cite{kingma2014adam}, frequently used to optimize optics in current state-of-the-art methodologies, are in direct contradiction to the HIL setup, where the hardware is a ‘black-box’ of unknown mathematical model and as a result nondifferentiable. Additionally, the efficiency of the CMA-ES strategy for Black-Box optimization of ISP is demonstrated by tuning a multi-objective highly nonlinear optimization problem in \cite{Mosleh_2020_CVPR}. \vspace{-0.8em}

\section{Experimental Results}
\begin{figure*}[t!]
	\centering
	\includegraphics[width=1\linewidth]{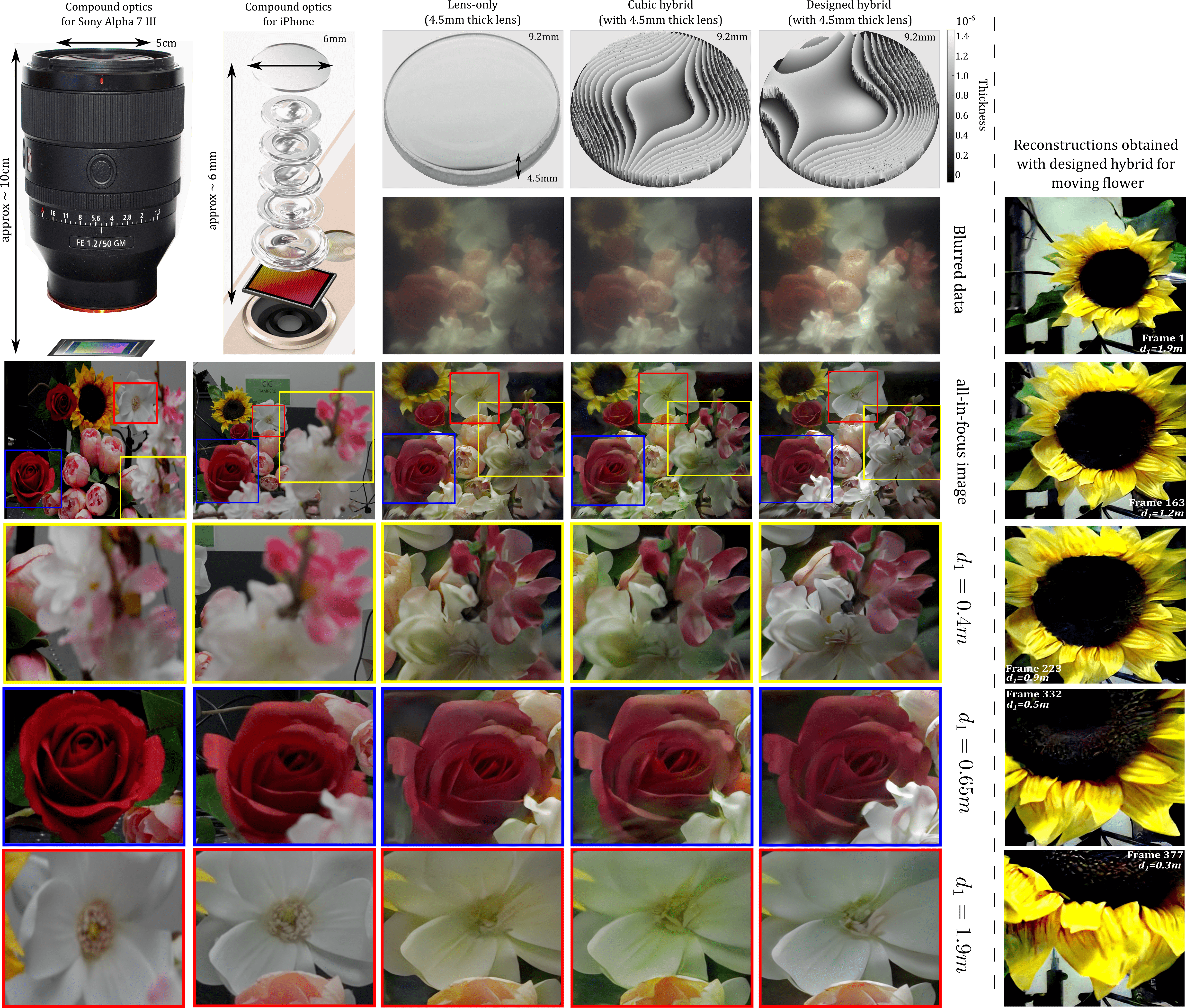}\vspace{-1em}
	\caption{\small Setup 2. Comparison of the diffractive optics versus the compound lens Sony A7 III (85mm focal length with F22) and iPhone Xs Max cameras. The obtained images are presented in row 3 with their enlarged fragments in rows 4, 5, 6 corresponding to three off-focus distances $d_{1}=0.4, 0.65, 1.9$~m, respectively. For comparison also included results obtained by lens-only and lens + SLM of cubic phase-pattern. The visual advantage in sharpness and color preservation is clearly in favor of the designed hybrid. The images of the sunflower moving in the $d_{1}$ depth range 1.9-0.3~m are shown in column~6. The sharpness and color presentation are very good despite variations of distance.} \vspace{-0.5em}
	\label{fig:recons}
\end{figure*}

In this section we evaluate and discuss the imaging results of the developed hybrid system over the testing dataset. These results are shown in Figures \ref{fig:system} and \ref{fig:recons}, respectively, for Setups 1 and 2 as noted in Figure \ref{fig:initial}. Remind that in Setup 1 the three monitors display images of the three distances from the sensor. In this scenario, we are able to evaluate the quality of reconstructions visually as well as numerically by PSNR values for each of the RGB color channels. Setup 2 is developed to evaluate EDoF where the scene is composed by flowers located at different distances from sensor to validate the performance of our design in real-world scenario. In this case, the quality of imaging can be evaluated only visually (qualitatively). 

\textit{Setup 1}. The results shown in Figure \ref{fig:system} are presented in three rows, respectively, for three depth distances (object-SLM): $d_1 = 0.5~m$ (row 1), $d_1 = 1.0~m$ (row 2), and $d_1 = 1.8$~m (row 3). The groups of columns \textit{Groundtruth}, \textit{Designed hybrid} and \textit{Lens-only} show the true images as displaced by the three monitors for different depths, reconstruction by the developed algorithm, and reconstruction by the lens-only system. In the last case, the SLM is switched off (no phase modulation) and imaging is produced by the lens only. For a fair comparison, the optimization of inverse imaging, in this case, is produced identically to those for the hybrid system. The experiments with the lens-only are produced in order to evaluate the effects of the phase-modulation by SLM. In the columns \textit{Designed hybrid} and \textit{Lens-only}, we can see blurred images used for reconstruction as well as the zoomed fragments of the reconstructions shown also for the true images. The zoomed sections for blurry and reconstructed images visually reveal clearly that the lens system suffers from strong chromatic aberrations and the quality of imaging is lost especially for defocus distances $d_1 = 0.5~m$ and $d_1 = 1.8~m$. The best performance for the both compared system is demonstrated for the focal distance $d_1 = 1.0~m$. However, the advantage of the hybrid is well seen in qualitative and numerical comparison. For the defocus distances the advantage of the designed hybrid is much more valuable. For instance, for $d_1 = 1.8~m$, the improvement of about 2dB is in favor of the hybrid. Moreover, for the designed hybrid, the PSNR values for different colors and depths are more or less the same at about 25~dB. It confirms that the designed hybrid imaging indeed demonstrates achromatic EDoF imaging.

\textit{Setup 2.} In this setup, the whole 3D scene is observed. It is the all-in-focus scenario when the goal is to get in-focus images for all object of different distances from SLM. Figure~\ref{fig:recons} shows the performance for: the designed hybrid (column 5), lens + cubic absolute phase (column 4), lens-only (column 3), and two compound commercial cameras, Sony A7 III (column 1) and iPhone Xs Max (column 2) cameras. In this experiment, 6 flowers are located at the different distances $d_1 = 0.4~m$ (cherry blossom), $0.65~m$ (first red rose), $1.0$~m (tulip flower), $1.6~m$ (second red rose), $1.8~m$ (sunflower), and $1.9$~m (magnolia flower) as it is shown in Figure \ref{fig:initial}. It is worth mentioning that for Sony and iPhone (compound optics), we adjusted the focusing point to $d_1 = 1.0~m$ (tulip flower) for a fair comparison with other imaging systems. For infocus point ($d_1 = 1$~m), Sony camera produces the best result but with a narrowed DoF. On the other hand, the imaging quality for iPhone camera is not good for distances less than 1.0~m.

The imaging results are presented in row 3. For more detailed comparison, the three zoom fragments of these images corresponding to different distances $d_1 $ are shown in rows 4, 5, and 6. The visual advantage of the hybrid in both the sharpness of imaging and the proper color preservation is quite obvious. The lens-only and cubic hybrid setups exhibit strong chromatic aberrations with less imaging accuracy for both focus and out-of-focus situations similar as it was seen in Figure \ref{fig:system}. The most important in these experiments is a comparison of our designed hybrid versus the compound lens Sony and iPhone cameras, where the designed hybrid clearly shows better all-in-focus imaging. In addition to these experiments, we acquired a video of the sunflower moving in the depth range (1.9-0.3)~m. A few frames of this video are presented in column 6 of Figure \ref{fig:recons}. One may note that despite of the distance variations the sunflower imaging is sharp with good color preservation for all distances. Note, that the shown frames are given for distances different from those used in optimization. In summary, these experiments suggest the effectiveness of the HIL-SLM setup as a base methodology for designing diffractive imaging systems. \vspace{-0.5em}

\section{Conclusion}
It is shown in this paper that the optimized hybrid optical system composed from refractive lens and DOE in the scenario of achromatic EDoF imaging demonstrates advanced performance as compared with: the single refractive lens, lens + cubic absolute phase component, and two compound commercial cameras, iPhone Xs Max and Sony A7 III cameras. In experiments, the hybrid optics is implemented by optical projection of the SLM phase pattern on a lens plane for the visible wavelength interval (400-700) nm and the depth-of-ﬁeld range (0.4-1.9)~m. Multiple results comparing the developed DOE imaging system with advanced conventional compound multi-lens cameras such as Sony A7 III and iPhone Xs Max cameras show its competitive imaging quality and advanced in all-in-focus sharp imaging.


\bibliographystyle{ACM-Reference-Format}
\bibliography{sample-bibliography}


\end{document}